\documentclass[
 reprint,
 nobibnotes,
 amsmath,amssymb,
 aps,
 pra,
]{revtex4-2}
\usepackage{graphicx}
\usepackage{dcolumn}
\usepackage{bm}
\usepackage[hidelinks]{hyperref}
\usepackage{caption}
\usepackage{subcaption}

\usepackage{color}

\begin{document}

\title{Optical Levitation of Arrays of Microspheres}

\author{Benjamin Siegel}
\author{Gadi Afek}
\altaffiliation{Present address: Quantum Art, Ltd., Nes-Ziona 7414003, Israel}
\author{Cecily Lowe}
\author{Jiaxiang Wang}
\author{Yu-Han Tseng}
\author{T. W. Penny}
\author{David C. Moore}
\affiliation{
Wright Laboratory, Department of Physics, Yale University, New Haven, Connecticut 06520, USA
}

\begin{abstract}

Levitated optomechanical systems are rapidly becoming leading tools for precision sensing of forces and accelerations acting on particles in the femtogram to nanogram mass range. These systems enable a high level of control over the sensor's center-of-mass motion, rotational degrees of freedom, and electric charge state. For many sensing applications, extending these techniques to arrays of sensors enables rejection of correlated noise sources and increases sensitivity to interactions that may be too rare or weak to detect with a single particle. Here we present techniques capable of trapping defect free, two-dimensional arrays of more than 25 microspheres in vacuum. These techniques provide independent control of the optical potential for each sphere. Simultaneous imaging of the motion of all spheres in the array is demonstrated using camera-based imaging, with optimized object tracking algorithms reaching a displacement sensitivity below 1~nm/$\sqrt{\mathrm{Hz}}$. Such arrays of levitated microspheres may find applications ranging from inertial sensing to searches for weakly interacting particles such as dark matter.

\end{abstract}

\maketitle

\section{\label{sec:intro}Introduction}

Levitated optomechanics provides a powerful tool for precision sensing of forces and accelerations acting on nanometer to micron-scale particles. Thermal and mechanical isolation of levitated particles in vacuum has enabled force sensitivities at the yoctoNewton level~\cite{liang_yoctonewton_2023}. Precise control of a levitated nanosphere's center-of-mass motion has been demonstrated by cooling one~\cite{delic_cooling_2020,magrini_real-time_2021,tebbenjohanns_quantum_2021,kamba_optical_2022} or more~\cite{ranfagni_two-dimensional_2022,piotrowski_simultaneous_2023} degrees of freedom to the quantum ground state. The high sensitivity and control of such systems can enable tests of quantum mechanics of macroscopic systems~\cite{romero-isart_quantum_2011,arndt_testing_2014,romero-isart_large_2011,romero-isart_optically_2011,weiss_large_2021} or searches for interactions from physics beyond the Standard Model of particle physics~\cite{Moore:2020QST_review,kawasaki_search_2019,carney_searches_2023,kilian_dark_2024}. 

Such systems have already been utilized to search for dark matter~\cite{monteiro_search_2020} and to detect the force imparted by a single nuclear decay through the mechanical recoil of the sensor~\cite{wang_mechanical_2024}. They have been proposed for high-frequency gravitational wave detection~\cite{aggarwal_searching_2022} and investigating quantum friction~\cite{manjavacas_vacuum_2010,zhao_rotational_2012}. For some of these applications, while a single sensor already has sufficient detection sensitivity, the small cross section or limited ability to reject correlated noise limits the experimental sensitivity. Scaling to an array of levitated sensors benefits applications where the increase in mass or number of particles enhances the ability to see rare events (e.g., by increasing the cross section for weakly interacting particles)~\cite{afek_coherent_2022,carney_searches_2023}, or small forces correlated among many sensors that could be detected below the noise floor of a single sensor~\cite{carney_ultralight_2021,brady_entangled_2022,li_search_2023,higgins_maglev_2024,amaral_first_2024}. An array with tunable interparticle interactions can also enable studies of non-Hermitian interactions~\cite{yokomizo_non-hermitian_2023} or possibly quantum entanglement of macroscopic systems~\cite{rudolph_entangling_2020,chauhan_tuneable_2022}.\par
In parallel to the development of neutral atom tweezer arrays for quantum computing~\cite{anderegg_optical_2019,scholl_quantum_2021,ma_universal_2022,kim_quantum_2024,bluvstein_logical_2024,pause_supercharged_2024,manetsch_tweezer_2024,menon_integrated_2024}, there has been significant recent work to realize arrays of optically trapped nanospheres in vacuum for these applications. Recent work has demonstrated coupling of spheres through Coulomb~\cite{deplano_coulomb_2024} and optical~\cite{arita_optical_2018,rieser_tunable_2022,reisenbauer_non-hermitian_2024,vijayan_cavity-mediated_2024,svak_stochastic_2021,wu_coupler_2024} interactions. Arrays of up to 9 nanospheres have been trapped~\cite{yan_-demand_2023}. Cold damping~\cite{liska_cold_2023,vijayan_scalable_2023} and sympathetic cooling~\cite{arita_all-optical_2022} was achieved for pairs of particles. Electromagnetic trapping has made similar advances, with demonstrations of sympathetic cooling between two spheres~\cite{penny_sympathetic_2023,bykov_3d_2023} and scalable detection and cooling using cameras~\cite{minowa_imaging_2022,ren_event-based_2022,ren_neuromorphic_2024} implemented using radio frequency Paul traps. Coulomb interactions between spheres co-trapped in a magnetic potential have also been studied~\cite{slezak_microsphere_2019}. In scaling these platforms to larger numbers of particles, possible challenges include independent control of the motion of a single sphere without affecting other spheres in the array, and sensing the center-of-mass motion degrees of freedom for an increasing number of traps while minimizing the required number of additional sensors and readout electronics.
\par
In this paper, we trap and monitor the motion of an array of 25 microspheres in vacuum by adapting techniques previously developed for fluid-based optical tweezer arrays. Using time-sharing~\cite{sasaki_laser-scanning_1991,sasaki_pattern_1991,visscher_micromanipulation_1993,visscher_construction_1996} of a single trapping beam, we demonstrate individual control of the traps, allowing for sorting and creation of arbitrary geometries of arrays. Several object tracking algorithms, including a convolutional neural network, are used to analyze camera images of an array to determine each sphere's motion individually. Unlike in fluid-based systems, closed-loop feedback based on each sphere's motion is typically required to maintain the stability of particles in vacuum-based microsphere traps and to cool their motion~\cite{monteiro_force_2020}. To demonstrate possible imaging methods suitable for implementing closed-loop feedback, multiplexing of multiple spheres' position signals on a single sensor is demonstrated using back focal plane imaging~\cite{gittes_interference_1998,pralle_three-dimensional_1999,farre_optimized_2012,martinez_back-focal-plane_2012,kurvits_comparative_2015,ruh_fast_2011}. With this multiplexing and time-sharing approach, we provide a method capable of independent control and imaging of each microsphere's motion that can ultimately be scaled to large arrays of nanogram mass particles.

\begin{figure*}
    \includegraphics[width=0.8\textwidth]{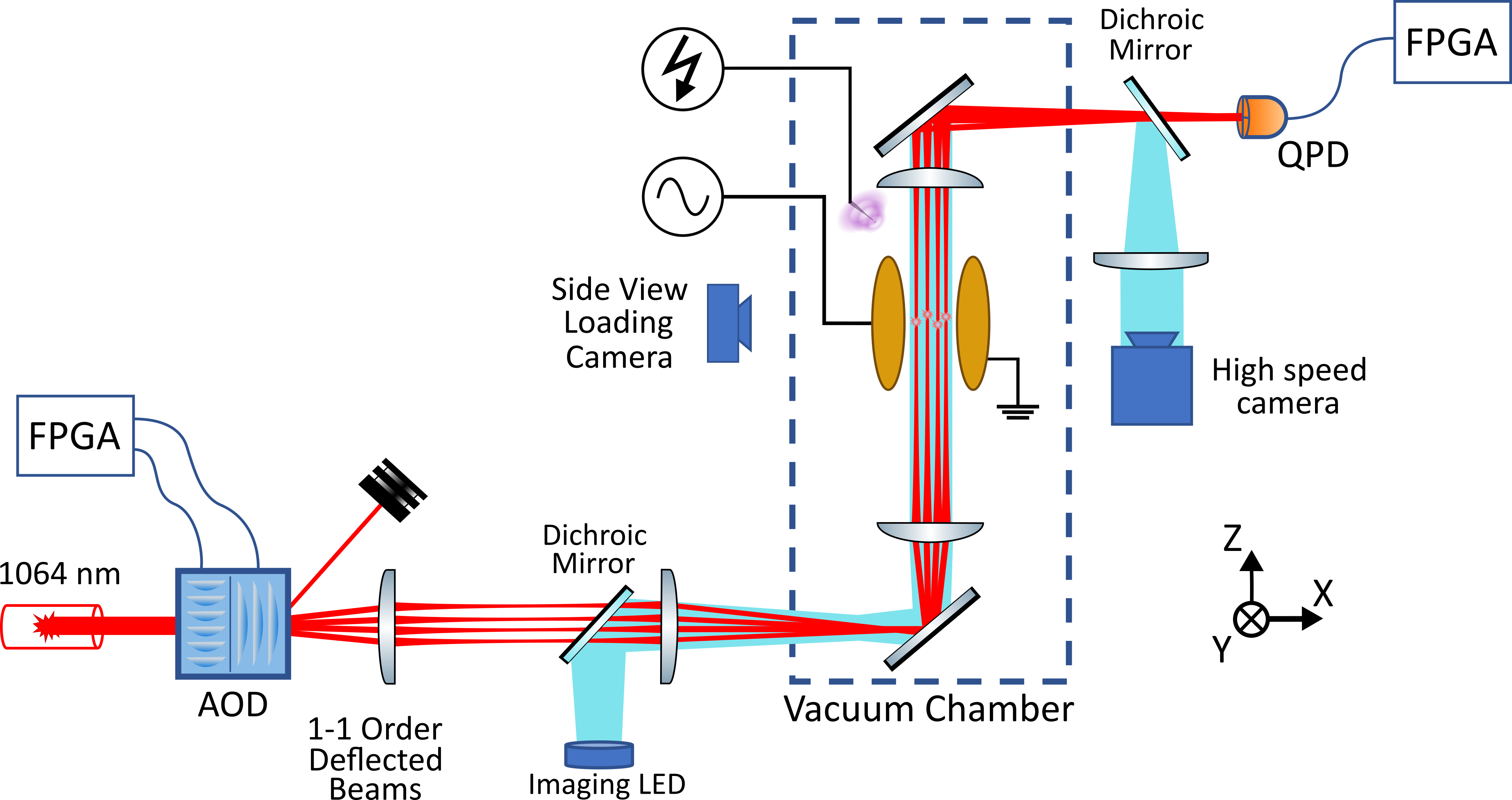}
    \caption{Simplified optical schematic. A two-dimensional acousto-optic deflector (AOD) splits the high powered 1064~nm input laser into individual trapping beams. The trapping beams are generated through a time-sharing approach with just a single element of the beam array deflected at any point in time (further discussed in section~\ref{sec:Timeshare}). This figure depicts the time averaged state of the array, displaying all the elements at once. Only the first order deflected elements of each crystal are utilized, with other orders discarded. These beams are then imaged into the vacuum chamber and vertically oriented, where they are focused with an aspheric lens with a focal length of 25~mm and numerical aperture (NA) of 0.025 to form gravito-optical traps with waists of approximately 10~$\mu$m. Using a dichroic mirror, collimated light from an LED with a wide beam diameter ( $\geq$ 4~mm) is co-aligned with the 1064~nm beams to image the spheres. The imaging and 1064~nm light are then collected by an objective lens with an NA of 0.25. The 1064 nm trapping light is collimated and measured with a quadrant photodiode, while a high-speed camera observes the image plane of the objective lens to monitor the spheres' motion via the imaging light. In the chamber two electrodes, positioned opposite from each other to surround the trapping region, can be used to determine the electric charge of the spheres by tracking the amplitude of their motion in response to an oscillating electric field. A needle connected to a DC high voltage supply is placed near the trapping region to generate plasma for the loading sequence (discussed in section \ref{sec:loading}). A second camera is positioned to monitor a side view of the trapping region during the loading process.}
    \label{fig:optics}
\end{figure*}

\section{\label{sec:methods}Methods}

The experimental configuration follows a similar design to previous experiments levitating single nanogram mass particles in vacuum ~\cite{ashkin_feedback_1977,ashkin_optical_1971,ashkin_optical_1976,moore_search_2014,blakemore_librational_2022,afek_control_2021,afek_limits_2021, monteiro_force_2020, monteiro_optical_2018}. A vertically oriented 1064~nm laser beam forms a trap capable of levitating spheres ranging from 1-30~$\mu$m in diameter~\cite{monteiro_optical_2018}. In this paper, silica spheres with an average diameter of $10.8\ \mu$m are trapped~\footnote{Purchased from Corpuscular, Inc. \url{http://www.microspheres-nanospheres.com/Microspheres/Inorganic/Silica/SiO2 20Plain.htm}}.

\begin{figure*}
    \centering
    \includegraphics{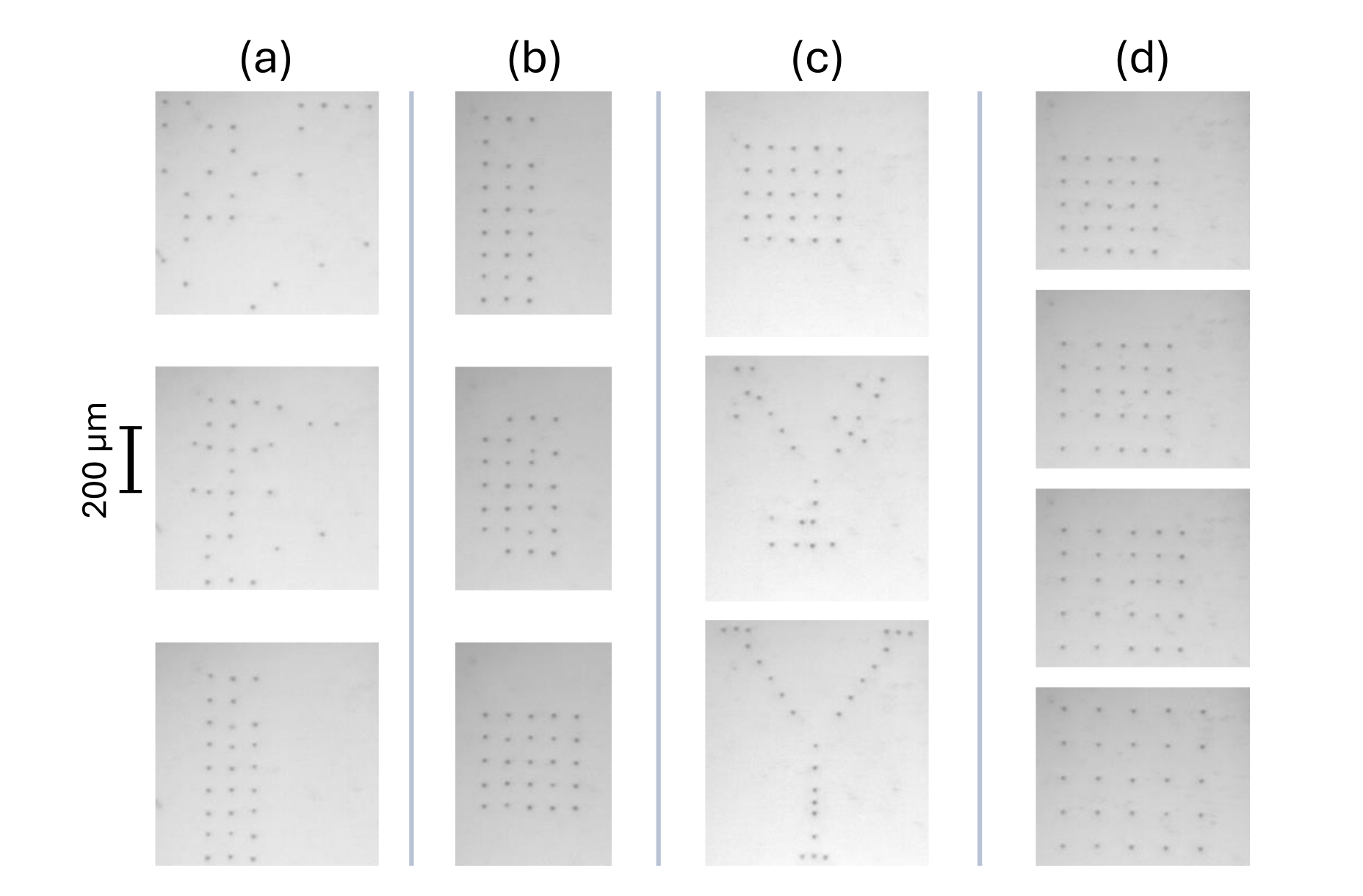}
    \caption{Photographs of trapped microsphere arrays in vacuum. Pictures in column (a) show a demonstration of rearranging an initially partially filled array following the loading process to a dense grid. Column (b) shows a second rearrangement of an imperfect grid to a fully filled grid with different dimensions. Column (c) gives an example of the independence of the traps enabling creation of arbitrary geometries. Column (d) illustrates the ability to change the intersphere spacing for the columns and rows. The scale bar applies to all images in the figure. Each sequence consists of snapshots from videos recording the evolution of the traps during the sorting process. The colors in all images are digitally enhanced to increase the contrast of the spheres relative to the background for visibility.
    }
    \label{fig:rearrange}
\end{figure*}

A simplified schematic of the system is shown in Fig.~\ref{fig:optics}. To create multiple traps, the laser propagates through an AA Opto-electronic DTSXY-400-1064 two-dimensional acousto-optic deflector (AOD), where radio frequency (RF) tones generated by a Field-Programmable Gate Array (FPGA) deflect the beam. The laser source is a IPG continuous wave laser (YLR-100-1064-LP) capable of producing 100~W. The laser power requirements are currently set by the procedure for loading the traps, which requires higher power (about 200~mW per trap) than for stable trapping. To account for losses from the diffraction efficiency of the AOD and in the optical system, a total power of 30~W is required to create an array of 100 traps. The array of beams is then imaged into the vacuum chamber, where their waists are focused to $\approx 10$~$\mu$m. The same 1064~nm beams used for trapping are also used for detection of the sphere position by imaging the transmitted light on a quadrant photodiode (QPD). Additionally, co-propagating incoherent green light from a light emitting diode (LED) is used for camera-based microscopy. These detection methods will be discussed in sections \ref{sec:BFP} and \ref{sec:camera} respectively.

\subsection{\label{sec:Timeshare}Time-Shared Generation of an Array}
To generate the individual trapping beams, we use a time-sharing based approach with a single laser~\cite{visscher_construction_1996,visscher_micromanipulation_1993,sasaki_laser-scanning_1991,sasaki_pattern_1991}. In this approach, the FPGA generates one frequency tone for each AOD axis at a time, forming one element of the array. The FPGA then cycles through pairs of tones to form each trap. With this method, there is always only one trap illuminated. By keeping the time spent per array site (dwell time) short enough, we reach the regime where the rate at which each site is addressed (cycle-rate) is much greater than the resonant frequencies of our spheres, forming pseudo-continuous traps. This time-shared approach is possible since the gravito-optically trapped microspheres have resonant frequencies for their center-of-mass motion of $\approx 30$~Hz for the axial direction and $\approx 100$~Hz for the radial directions. These frequencies enable the creation of arrays up to approximately 10$\times$10 elements using this method without loss of the microspheres due to the cycling. Since only a single sphere is illuminated at any given time, this time-sharing method enables fully independent control of the position and power of the optical trap at each site in the array. Utilizing this feature, we can create arbitrary two-dimensional geometries for the traps, as shown in Fig.~\ref{fig:rearrange}. For monitoring the position of the spheres in the array, time-sharing also enables the signals from each site to be distinguished since they are temporally separated. The signals can then be multiplexed onto a single sensor, as discussed further in section~\ref{sec:BFP}. These features enable us to realize individual feedback on each sphere in future work.
\par

\begin{figure*}[t]
    \includegraphics[width=0.98\linewidth]{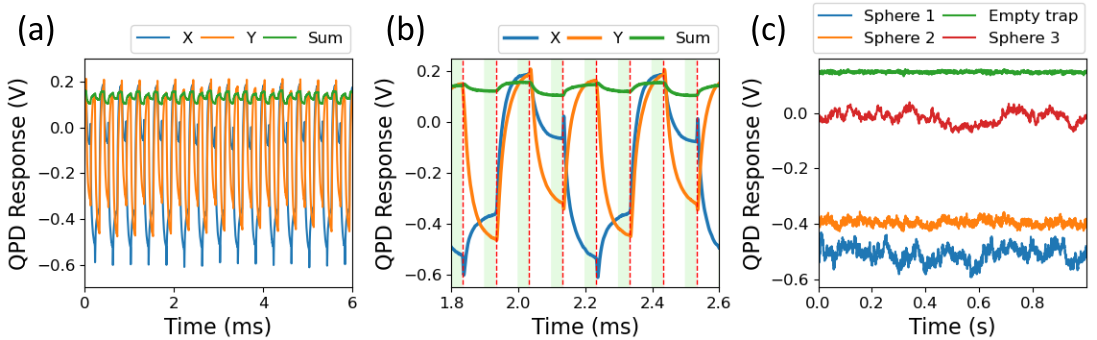}
    \caption{Example of data taken with the quadrant photodiode using back focal plane imaging. (a) Unprocessed signals from the QPD's \textit{X}, \textit{Y}, and sum channels. (b) Illustration of the method used to separate the different traps signals. The data is divided into segments by vertical dashed red lines showing when the timeshare algorithm switches to the next trap. In this data, we are cycling through four traps, as shown by the repetition of four distinct segments, and three are populated with spheres. Due to the long transient from the step between traps, only the shaded light green regions are used in determining a trap's position measurement. Each segment is then averaged and taken to be one data point for its respective trap. (c) Time-stream of the QPD's position measurement for the three trapped microspheres as well as the empty trap, generated by stitching each data points from the time windows shown in (b).}
    \label{fig:qpd}
\end{figure*}

Alternative methods where a comb of frequencies driving each AOD is used to generate an array of traps were also considered. In these methods, the configuration is limited to rectangular grids, with all the elements in a column or row coupled. Such coupling would complicate modulating the optical potential seen by a single sphere, since all the spheres in the same row and column would be modulated as well. To have independent traps using a continuously illuminated array, only the diagonal sites from the AOD's output could be used, limiting the number of spheres that could be simultaneously trapped for a given total laser power. Arrays generated by a comb of frequencies can be multiplexed onto a single sensor as well~\cite{vijayan_scalable_2023}, but require more complex and computationally expensive FPGA firmware to enable demodulation of signals separated by frequency offsets in the MHz range. However, the primary advantage over the time-sharing approach is that the cycle-rate will eventually introduce noticeable micromotion and limit the array size (for arrays $\gtrsim 100$ microspheres). Future implementations of very large rectangular arrays may require such frequency-based multiplexing.

\par

For loading a 100 beam array with the time-sharing method, we cycle through the tones, dwelling on each for 20~$\mu$s, for a full cycle period of 2~ms. Experimentally, we have seen this long dwell time leads to more efficient loading of the array, even though each sphere is addressed only at a cycle-rate of 500~Hz. After loading, the minimum dwell time is ultimately limited by the access time of the AOD and laser beam width at its input. With the minimum beam width possible while keeping the laser intensity below the AOD's damage threshold, dwell times of $\approx6.2$~$\mu$s per tone are achievable without distorting the outgoing beam of the AOD. For arrays as large as 100 traps, each site in the array can be addressed at cycle-rates greater than 1.5~kHz. Since this cycle-rate is much larger than the resonant frequency, this minimizes any driven motion of the spheres arising from the trap cycling.

\subsection{\label{sec:BFP}Multiplexed Readout of an Array}
To minimize the number of sensors required for an array of trapped microspheres, we use back focal plane imaging~\cite{ghislain_scanning-force_1993,ghislain_measurement_1994,pralle_three-dimensional_1999,farre_optimized_2012,martinez_back-focal-plane_2012,kurvits_comparative_2015,wiley-vch_verlag_gmbh__co_kgaa_fourier_2019}. This is a well-established technique in fluid optical tweezers, introduced by Gittes and Schmidt~\cite{gittes_interference_1998}. A QPD is placed in the Fourier plane of our objective lens, where it measures the interference pattern of the light passing through the microspheres when combined with the undeflected light from the 1064~nm trapping beams. Since the array consists of parallel beams in the trapping region where the beams are focused, the beams converge into one spot in the Fourier plane, and a single QPD is sufficient to detect the motion of all the spheres.\par

The time-shared trap generation also enables a simple FGPA algorithm to differentiate the measurements of each sphere's position using the QPD. Since the readout of each sphere's position is temporally separated, each time window corresponding to the illumination of a given sphere can be separately identified after calibration of the delay between generation of a given tone and the QPD response. In contrast to a frequency multiplexed approach, no high-bandwidth frequency-domain channelization is required to be implemented in real time on the FPGA.

When time sharing the array such that the cycle-rate is much greater than the resonant frequencies of the spheres' motion, the average signal measured during each dwell time can be treated as a single measurement of a given sphere's position. Since the sphere's motion is highly attenuated at frequencies well above its resonant frequency, the variation of the sphere position in one dwell time is negligible. Thus, the FPGA averages the signal over the dwell time for a single site and downsamples this data in time to record a single datapoint for the position of the sphere. These points are then used to form the measured position time series for each sphere.

An example of data taken with this method is shown in Fig. \ref{fig:qpd}. To mitigate effects of the currently used QPD's slew rate, the initial portion of the signal during each dwell time is ignored, and only data within the light green regions in Fig.~\ref{fig:qpd}b is averaged. An example reconstructed time series of the positions of multiple spheres is shown in Fig.~\ref{fig:qpd}c. While these measurements demonstrate the principle of a time-multiplexed position measurement of multiple spheres using a single QPD, future work is required to implement higher-bandwidth electronics for the QPD readout to minimize loss of data and systematic errors arising from the relatively slow rise time of the QPD readout electronics used here.

\subsection{\label{sec:camera}Camera Microscopy and Object Tracking through Video Analysis}
In addition to the QPD, we use a high-speed camera (GigaSens HS 2-2247-M-CXP2) with an acquisition rate up to 2247 frames per seconds (fps) at full resolution, to directly image the spheres, providing a secondary detection scheme that can be used in offline reconstruction of the sphere positions. Recent work using camera based detection of levitated particles has demonstrated high-resolution position detection~\cite{lewandowski_high-sensitivity_2021}, position reconstruction limited only by data transfer speeds~\cite{lansdorp_high-speed_2013}, and cooling of trapped particles with real time feedback~\cite{minowa_imaging_2022,ren_neuromorphic_2024}. For the system presented here, the camera provides a more accurate method for monitoring the positions of an array of spheres due to the non-negligible rise time of the QPD readout electronics compared to the time-sharing dwell time, although with more optimized QPD readout electronics we expect the QPD to reach equivalent, or better, displacement sensitivity.

\begin{figure}
    \centering
    \includegraphics[width=\linewidth]{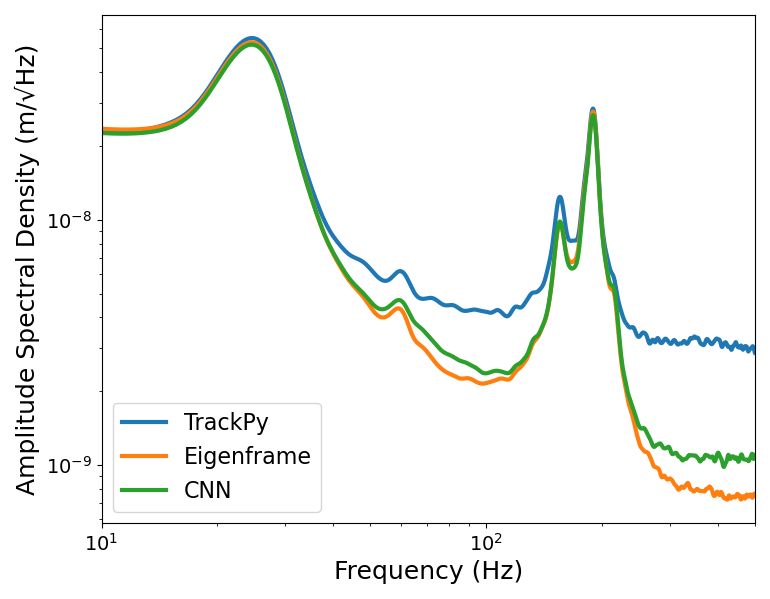}
    \caption{Reconstruction of the amplitude spectral density using the various position reconstruction algorithms described in the text. All reconstruction algorithms agree in the frequency range near the resonant peaks. However, as inferred from the noise level at high frequencies, where the sphere's motion is highly suppressed, the CNN and eigenframe methods achieve better displacement sensitivities than the centroid tracking approach. The amplitude spectral densities are taken from the \textit{X} motion of a single sphere trapped at 0.2 mbar over the course of a 20 seconds long, 1000 fps video.}
    \label{fig:tracker_comparison}
\end{figure}

For camera imaging, the 1064~nm trapping light is blocked and the imaging LED is used for bright-field microscopy. Three position reconstruction methods were studied: a centroid object tracker~\cite{crocker_methods_1996} package developed for Python (\texttt{TrackPy}~\cite{allan_soft-mattertrackpy_2024}), a template based algorithm employing an eigenframe translation analysis~\cite{lewandowski_high-sensitivity_2021}, and a convolutional neural network (CNN)~\cite{carleo_machine_2019,karagiorgi_machine_2022}.

While the first two methods directly employ algorithms developed in previous work, the CNN-based method was developed and optimized specifically for analyzing images of the arrays taken in this work. The CNN is constructed using a two-layer architecture for particle localization, with tunable hyperparameters optimized via a grid search. To train the model, a simulated dataset is generated by superimposing a synthetic circular particle onto experimental background images. Each simulated image is shifted by random horizontal and vertical offsets drawn from a normal distribution, which are recorded as truth labels for supervised learning. The CNN is trained on this data to predict particle positions corresponding to these shifts. Performance is evaluated using the Mean Squared Error (MSE) loss function to quantify the model's predictive accuracy on a validation set derived from the same simulated dataset. The measured error in the validation set was $4.9 \times 10^{-4}$ pixels, which demonstrates that sub-pixel accuracy of the model is achieved in particle position estimation.

Fig.~\ref{fig:tracker_comparison} compares the performance of these methods in reconstructing the motion of a single trapped sphere. While all methods are able to successfully reconstruct the sphere motion near its resonant frequency, the eigenframe method and CNN reach substantially lower noise levels at high frequencies. At these frequencies, the motion of the sphere is highly suppressed, allowing the displacement sensitivity for a given reconstruction algorithm to be determined. Both the eigenframe method and CNN reach displacement sensitivities $\lesssim1$~nm/$\sqrt{\mathrm{Hz}}$, which are substantially better than the centroid tracking algorithm. While the eigenframe method slightly outperforms the CNN in this implementation, once the network is trained, the computation time for the CNN is substantially lower than for the eigenframe method.

\begin{figure*}[t]
\centering
    \begin{subfigure}[t]{0.3\textwidth}
        \includegraphics[width=\textwidth]{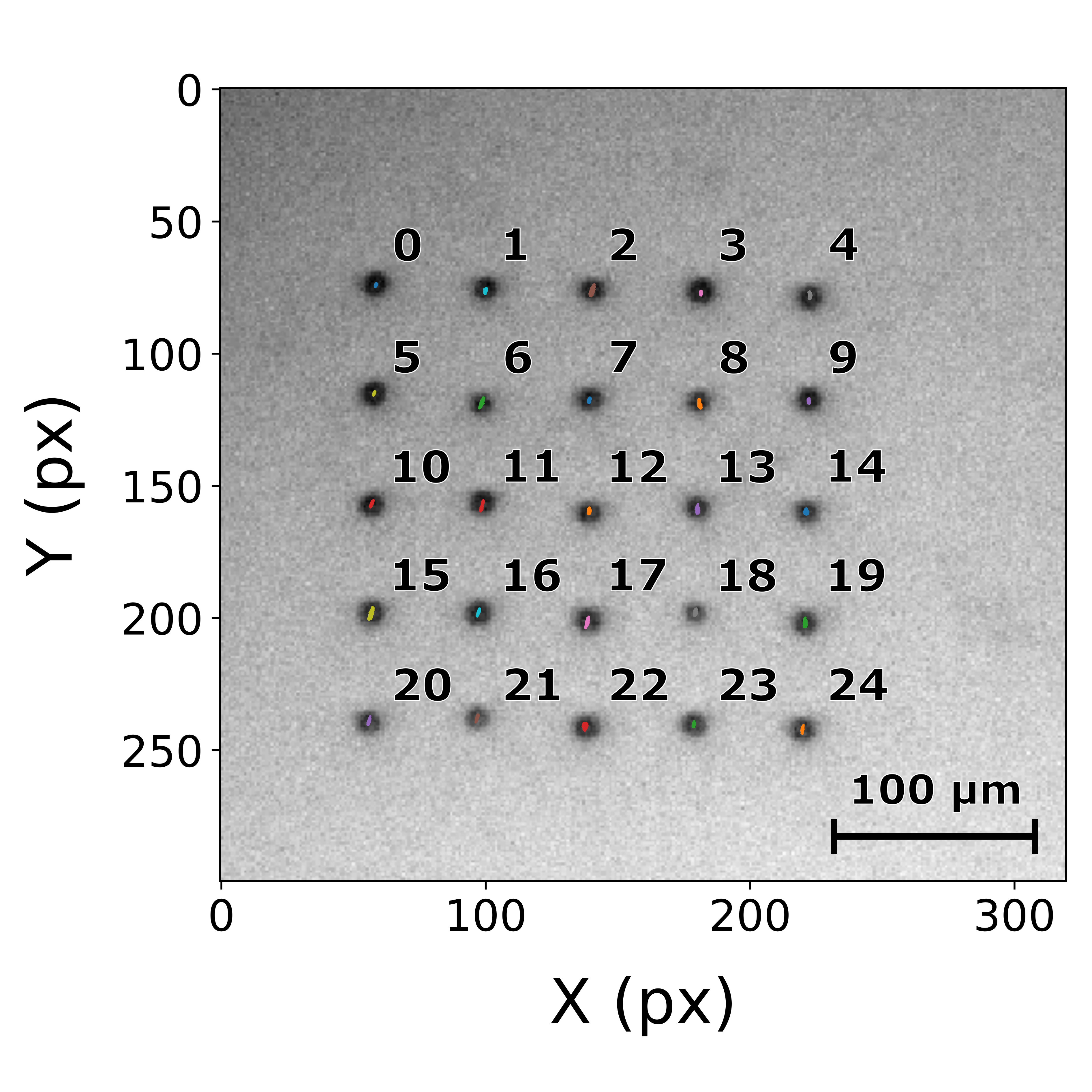}
        \caption{Traces of spheres' positions over time}
        \label{fig:index}
    \end{subfigure}
    \begin{subfigure}[t]{0.3\textwidth}
        \includegraphics[width=\textwidth]{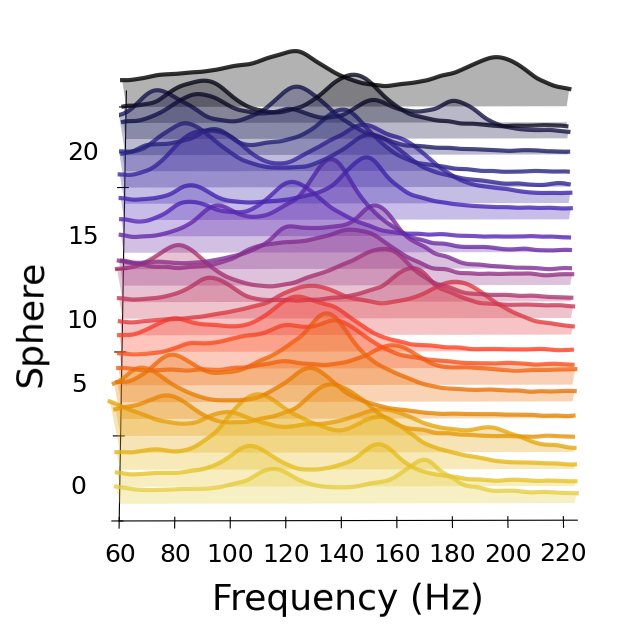}
        \caption{Spectra of \textit{X} motion}
        \label{fig:xlargearray}
    \end{subfigure}
    \begin{subfigure}[t]{0.3\textwidth}
        \includegraphics[width=\textwidth]{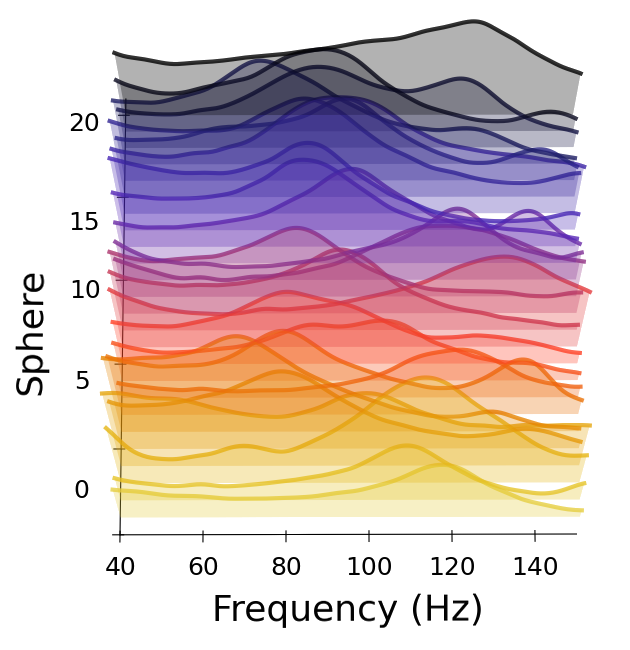}
        \caption{Spectra of \textit{Y} motion}
        \label{fig:ylargearray}
    \end{subfigure}
\caption{(a) Motion of 25 spheres trapped at 0.4~mbar in a 5$\times$5 array with $57\pm0.2~\mu$m spacing between the rows and columns of traps. The axes of the plot are in units of the camera's pixels (px), with the colored tracks (smaller than the sphere size) showing the spheres' position evolution over the course of a video. The plot is overlaid on a frame of the video with a scaling bar showing distance in $\mu$m. The right two figures show the $X$ (b) and $Y$ (c) normalized amplitude spectral densities reconstructed for each of the 25 spheres from a 1000~fps, 24~s long video recorded with the high-speed camera. The vertical axis is the index of the sphere, as labeled in (a). The amplitude spectral densities of the spheres are scaled and offset for visibility.}
\label{fig:largearray}
\end{figure*}

\subsection{\label{sec:loading}Loading Techniques}
Microspheres are loaded into the array using a piezoelectric transducer to detach spheres from a glass surface positioned above the trap~\cite{ashkin_optical_1971,Tongcang_thesis,weisman_apparatus_2022,blakemore_librational_2022,monteiro_force_2020}. The spheres then fall through a mm sized pinhole above the array and randomly fill the traps if they pass sufficiently close to a given trap focus. This loading procedure is performed at pressures of $\approx3$~mbar to provide enough damping to slow the spheres after their release. To achieve high filling fractions when loading an array, high voltage is applied to a needle electrode in the trapping chamber, creating a plasma through corona discharge that electrically neutralizes the spheres as they fall~\cite{frimmer_controlling_2017,ricci_accurate_2019}. Triboelectric charging of the spheres provides a net charge of $10^3$ to $10^4$ excess electrons after being detached from the cantilever, which if not neutralized by the plasma causes sphere-to-sphere Coulomb interactions to overwhelm the optical forces of the trap. By decreasing the Coulomb interactions via corona discharge, the loading efficiency is increased from $\lesssim 5$\% in a complete loading sequence, to achieving over a third filling of the array.\par

To reach even higher loading fractions, an auxiliary trapping beam will be used to first trap and discharge spheres in a separate loading area, and then transfer these neutralized spheres into the array. We have demonstrated successful trapping in a beam that is separated from the 1064~nm trapping laser before the AOD, and has a greater field of motion inside the chamber than the array beams. To load additional spheres, the array is moved away from the loading region and the auxiliary beam is positioned under the cantilever. Once a sphere is trapped in this beam, it is moved via a motor controlled mirror to overlap with one of the array beams. Then, the amplitude of the auxiliary beam is attenuated until the sphere is trapped solely by the array's beam. Using this technique, we have demonstrated successful transfer of spheres between the trapping and array beam with near unity efficiency.

\subsection{\label{sec:sorting}Low Loss Organizing Algorithm}
Sorting algorithms to create defect free arrays have been an important technique in optical tweezer arrays for atoms~\cite{kim_situ_2016,endres_atom-by-atom_2016,barredo_atom-by-atom_2016,barredo_synthetic_2018}. Defect free arrays of microspheres are also advantageous to minimize unused traps and unnecessary laser power in the final array of traps, as well as in many sensing applications~\cite{afek_coherent_2022, Moore:2020QST_review}. The time-shared array generation developed here naturally allows for independent control of the traps.

\begin{figure*}[t]
    \centering
    \includegraphics[width=\textwidth]{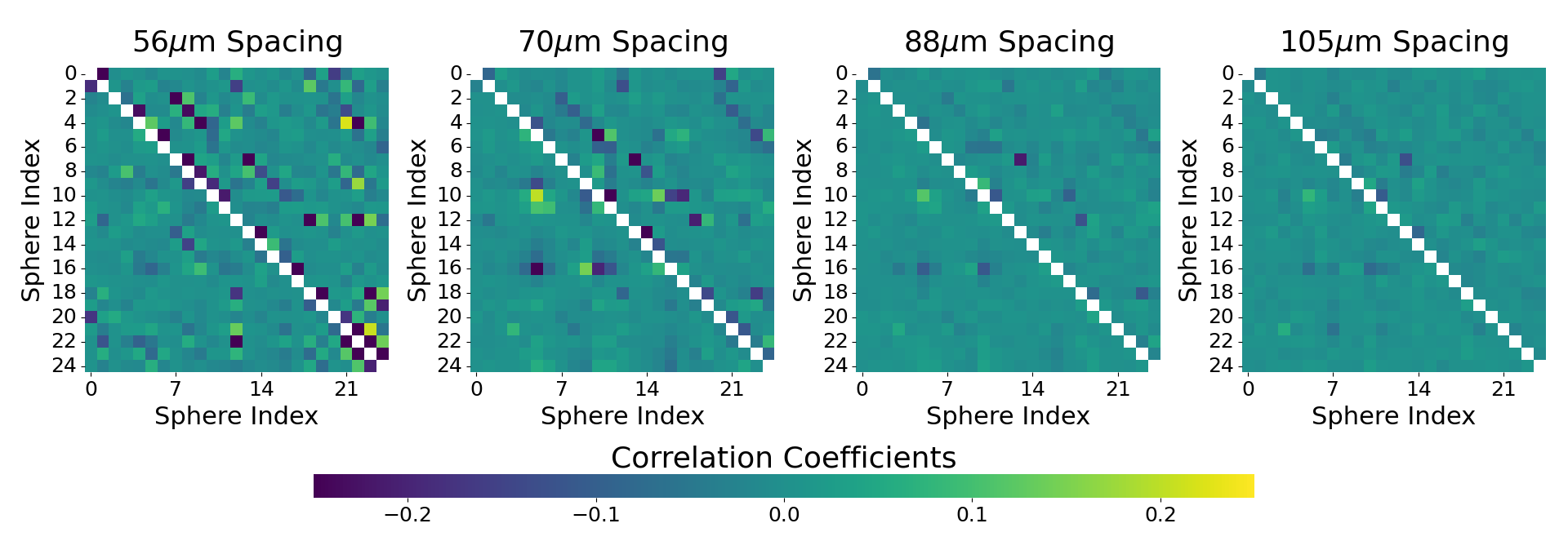}
\caption{Correlations of the motion of 25 spheres in a 5$\times$5 grid measured with different spacings between rows and columns. The array is indexed as shown in Fig. \ref{fig:index} The colormap depicts the Pearson correlation coefficient matrix of all the spheres' motion. The lower triangle of each plot represents the correlation of the \textit{X} motion between spheres, while the upper triangle represents the correlation of the \textit{Y} motion. The diagonal is intentionally left blank.}
\label{fig:corrs}
\end{figure*}

Offline code was developed to optimally sort arrays and apply the required movements using the AOD. Given a partially filled array, the initially filled spots are first matched with those for a densely packed array of the same number of elements. To minimize the total movement required for all the spheres, the Hungarian algorithm is implemented to solve this assignment problem~\cite{kuhn_hungarian_1955}. Direct paths are then drawn between the initial and final positions of each sphere. To maintain a minimum inter-sphere distance to prevent loss from collisions, the closeness of all paths is calculated for each step. If two paths create a collision point, the end points for the spheres are swapped, and the paths are recalculated and checked for collisions. This iterative calculation is continued until no more improvements can be made through swapping end points. If any collisions remain, we prevent them by delaying one of the organizing trajectories in time, holding the sphere in place while the other sphere can move through its path unimpeded.

When satisfactory paths are calculated, the AOD's control frequencies are mapped over these paths at a sufficiently slow speed ($20~\mu$m/s) to move the spheres without loss. Demonstrations of this algorithm to arrange arrays of different sizes and spacings are shown in Fig.~\ref{fig:rearrange}. This sorting algorithm is not limited to creating dense arrays, and can be used to create arbitrary 2D geometries with the trapped spheres, as shown in column (c) of Fig.~\ref{fig:rearrange}. 

\section{Simultaneous Measurement of Sphere Positions in a 25 Element Array}
Using the loading, organizing, and camera monitoring methods described above, we can monitor the motion of each sphere in a twenty five element array. Videos of the array's motion are recorded at 1000~fps, and offline reconstruction of the motion of each sphere within the array is performed using the algorithms described in Sec.~\ref{sec:camera}. Fig.~\ref{fig:largearray} shows the shape of the amplitude spectral densities of the motion in the \textit{X} and \textit{Y} directions for each sphere in a 5$\times$5 array trapped at 0.4~mbar. The rows and columns of this array are separated by $57\pm0.2~\mu$m. Spheres are indexed beginning at the top left of the array and increasing in each row from left to right. The difference in spectra between spheres can arise from residual differences in the electric charges after loading (and corresponding coupling to neighboring spheres) as well as variation in the sphere mass.

After loading, the dominant contributions to interactions between sphere are expected to arise from small overlaps in the tails of neighboring trapping beams and from Coulomb forces from the excess charge residing on the spheres after loading. While corona discharging is used here to reduce the sphere charges during loading, they typically retain a charge $\gtrsim$100~$e^{-}$. Future work will allow precise discharging of each sphere using ultra-violet light focused on each sphere individually~\cite{afek_limits_2021}. 

\par
To quantify the residual coupling between spheres, the Pearson correlation coefficients between the reconstructed sphere positions for this data are shown in Fig.~\ref{fig:corrs} as a function of array spacing. While the uncertainty on the electric charge of each microsphere prevents a quantitative simulation of the observed motion, correlated motion between nearest neighbors, across diagonal neighbors, and even correlations with spheres more than one column/row away is evident, with increasing strength at the smallest separations. However, at spacings $>$100~$\mu$m, these correlations are minimized, indicating that sensing applications that benefit from minimal coupling between spheres can be realized by minimizing the electric charge of the spheres and maximizing their spacing.

\section{Summary and Discussion}
We have presented a scalable method for optically levitating arrays of microspheres in vacuum by employing a time-sharing technique for array generation. This technique enables independent control of each trap in the array, which we utilize to create densely packed arrays and arrays of arbitrary configurations. By using back focal plane imaging, a single quadrant photodiode can be used to monitor the position of all spheres in the array, which in future work will be used to enable closed-loop feedback on the motion of the spheres. 

In the current system, we have also demonstrated monitoring of the motion of multiple trapped spheres using a high speed camera and reconstruction of the spheres' positions based on established algorithms, as well as a new CNN approach developed here. We demonstrate evidence of Coulomb coupling between the motion of neighboring spheres in the array, which can be minimized by increasing the array spacing and minimizing the electric charge of the spheres. In future work, site-addressable discharging with single electron charge precision will be implemented using a UV laser. These extensions will enable arrays of up to 10$\times$10 microspheres to be trapped in high vacuum, which is expected to substantially enhance the sensitivity of levitated optomechanical systems for applications ranging from inertial sensing to fundamental physics.

\begin{acknowledgments}
This work was supported by ONR Grant N00014-23-1-2600 and NSF Grant PHY-2109329. This material is based in part upon work supported by Alfred P. Sloan Foundation under Grant No. G-2023-21130, and is funded in part by the Gordon and Betty Moore Foundation through Grant GBMF12328, DOI 10.37807/GBMF12328. Yu-Han Tseng is supported by the Graduate Instrumentation Research Award (GIRA).
\end{acknowledgments}

\bibliography{manualbib.bib,papersources.bib}

\end{document}